\documentclass[aps,prl,twocolumn,floatfix,amsmath,amssymb,amsfonts,citeautoscript,longbibliography]{revtex4-2}
\usepackage[colorlinks=true,citecolor=blue]{hyperref}
\usepackage[all]{hypcap}
\usepackage{graphicx}
\newcommand{\revision}[1]{{{#1}}}
\newcommand{\revisiontwo}[1]{{{#1}}}

\begin{document}
\author{Mrinmoyee Saha}
\affiliation{Department of Applied Physics, Aalto University, 00076 Aalto, Finland}
\author{Luca Horray}
\affiliation{Department of Applied Physics, Aalto University, 00076 Aalto, Finland}
\author{Pedro Portugal}
\affiliation{Department of Applied Physics, Aalto University, 00076 Aalto, Finland}
\author{Christian Flindt}
\affiliation{Department of Applied Physics, Aalto University, 00076 Aalto, Finland}

\title{Negative currents in Fabry-Pérot cavities are caused by interfering paths}
    
\begin{abstract}
\revision{It has been predicted that the time-dependent current in a Fabry-Pérot cavity can turn negative even if the applied voltage pulses are always positive. It has also been suggested that the negative currents are related to interfering paths, however, an analytic description of this surprising phenomenon has so far been missing. \revisiontwo{Here we make use of Floquet scattering theory to demonstrate that the negative currents indeed are caused by the interference of scattering paths through the cavity with different numbers of roundtrips.} Our analytic result can be tested in future experiments as it predicts exactly how the effect should be washed out by an increasing temperature.} We show that a similar phenomenon is expected for the heat current, which may also turn negative.
\end{abstract}

\maketitle
		
\emph{Introduction.} Gigahertz voltage pulses have paved the way for experiments in high-frequency quantum transport, where just a single or a few electronic excitations are emitted into a mesoscopic conductor~\cite{Bocquillon:2014,Splettstoesser:2017,Bauerle:2018,acciai:2025}. Lorentzian pulses generate clean single-particle excitations known as levitons, which can be controlled and manipulated as photons in quantum optics~\cite{Levitov:1996,Levitov:1997,Keeling:2006,Dubois:2013,Jullien:2014,Assouline:2023,Chakraborti2025}. 
When emitted close to the Fermi level, these  excitations interact only weakly. By contrast, recent collision experiments have revealed strong interactions between high-energy excitations that arrive simultaneously on each side of a beam splitter~\cite{Wang:2023,Fletcher:2023,Ubbelohde:2023}. Moreover, by increasing the driving frequency and shortening the pulse duration, many dynamical phenomena may soon be discovered~\cite{Aluffi:2023,Ouacel:2025}.

At low temperatures, electronic excitations can experience quantum coherence over an entire mesoscopic structure, such that interference effects become observable. A prominent example is the Hong-Ou-Mandel effect~\cite{Hong:1987,Bocquillon:2013}, which recently was realized with electrons~\cite{Dubois:2013,Jullien:2014,Assouline:2023,Chakraborti2025}. In addition, Mach-Zehnder interferometers and Fabry-Pérot cavities have been implemented with mesoscopic circuits~\cite{Liang_2001,Ji:2003,Ofek:2010}. Such devices have mainly been explored using static voltages, however, recently, a Fabry-Pérot cavity driven by single-electron pulses was used for time-resolved sensing of electromagnetic
fields~\cite{Bartolomei:2025}. Moreover, several theoretical predictions have been made for interferometers driven by voltage pulses~\cite{Bocquillon:2014,Splettstoesser:2017,Bauerle:2018,Burset:2019,Kotilahti:2021}. For instance, \revision{using advanced numerical simulations, Gaury and Waintal have shown that the current in a Fabry-Pérot cavity may turn negative, although the time-dependent voltage is always positive, and it has been suggested that the negative currents are caused by interfering paths~\cite{Gaury:2014,Gaury:2014b}}. \revisiontwo{This surprising phenomenon may soon be within experimental reach, and it is therefore a timely task to develop an analytic description, which can be tested in future experiments.}

\begin{figure}[b!]
    \centering  
    \includegraphics[width = 0.99\columnwidth]{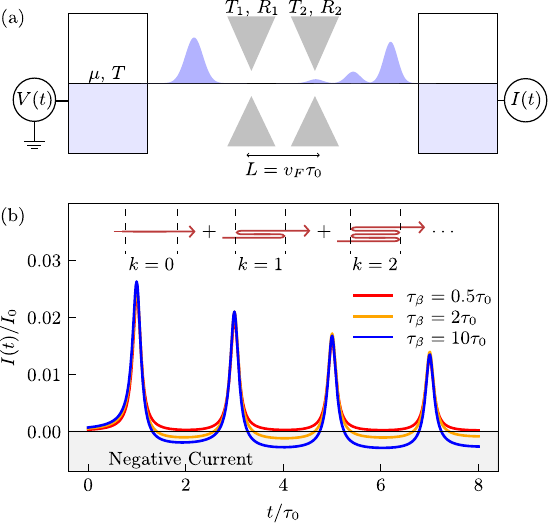}
    \caption{Negative currents in a Fabry-P\' erot~cavity. (a) The device consists of two quantum point contacts with transmission and reflection probabilities \revision{$T_{1,2}$ and $R_{1,2}=1-T_{1,2}$}, separated by the distance $L$. Electrons travel at the Fermi velocity~$v_F$. (b)~Time-dependent electric current in response to a Lorentzian voltage pulse of halfwidth $\Gamma = 0.1\tau_0$, which excites an average charge of $q=0.75$ electrons. \revision{In Figs.~\ref{fig:setup}-\ref{fig:heatcurrent}, the reflection probabilities are~$R_{1,2}=0.9$}, and we have defined $I_0 = \mathrm{e}/\tau_0$. We show results for different temperatures given by the thermal timescale~$\tau_\beta= \hbar/\pi k_BT$. The transmission amplitude in Eq.~(\ref{eq:SFabry}) is the sum over all scattering paths from the source to the drain involving $k=0,1,2,\ldots$ round~trips.}
    \label{fig:setup}
\end{figure}

In this Letter, \revision{we present an analytic theory of negative currents} in mesoscopic Fabry-Pérot cavities driven by voltage pulses as illustrated in Fig.~\ref{fig:setup}(a). 
To this end, we make use of Floquet scattering theory to show that the negative currents \revision{indeed} are caused by interferences between scattering paths through the cavity with different numbers of roundtrips; \revision{somewhat similarly to how quantum interference can lead to negative probabilities in full counting statistics~\cite{Hofer:2016}. Our theory can be tested in future experiments, since it predicts how the effect should be washed out with increasing temperature as shown in Fig.~\ref{fig:setup}(b).} We extend our analysis to the heat transport by showing that the heat current can also turn negative. 

\emph{Fabry-Pérot cavity.} Figure~\ref{fig:setup}(a) shows an electronic Fabry-Pérot cavity made up of two quantum point contacts that are separated by the distance $L$. We can write the transmission amplitude of the device as a sum over all possible scattering paths through the cavity,
\begin{equation}
	\label{eq:SFabry}
        t(E) =t_1t_2 e^{iE\tau_0/\hbar} \sum_{k=0}^\infty \left[r_1r_2 e^{i2 E\tau_0/\hbar}\right]^k, 
\end{equation}
where $\tau_0=L/v_F$ is the time it takes to travel between the quantum point contacts, with \revision{(real)} transmission and reflection amplitudes  $t_{1,2}$ and $r_{1,2}$, at the Fermi velocity~$v_F$~\cite{Lesovik:2011}.  With a constant voltage, the current through the interferometer is determined by the transmission function $T(E)=|t(E)|^2$, \revision{and it is constant.} By contrast, if driven by voltage pulses, the current depends on the transmission amplitude at two different energies, and there will be products of amplitudes corresponding to paths with different numbers of roundtrips, leading to interferences and negative currents as we will see.

We now consider voltage pulses that are applied to the source, while the time-dependent current is measured in the drain. To be specific, we consider Lorentzian voltage pulses, but one can also analyze other time-dependent voltages. The sequence of Lorentzian pulses reads
\begin{equation}
\mathrm{e}V(t) = \frac{q\hbar\Omega}{\pi}\sum_{n}\frac{\Gamma \mathcal T}{\Gamma^2+(t-n\mathcal T)^2},
\label{eq:levitons}
\end{equation}
where $\Omega=2\pi/\mathcal{T}$ 
is the driving frequency, and $\mathcal T$ is the period. The halfwidth of the pulses is denoted by $\Gamma$, and $q$ is the average charge excited by each pulse. To evaluate the electric current, we make use of Floquet scattering theory and thus consider the scattering phase due to the voltage, $\mathcal J(t)=\exp\{-i\phi(t)\}$ with $\phi(t)=\mathrm{e}\int_{-\infty}^t dt' V(t')/\hbar$~\cite{Moskalets_book,Dubois:2013b,Brandner:2020}.
Its Fourier components, $\mathcal J_n=\int_0^\mathcal{T}dt \mathcal J(t)e^{in\Omega t}/\mathcal{T}$, are the amplitudes for an electron in the source to change its energy from $E$ to $E_n=E+n\hbar\Omega$ by exchanging $n$ modulation quanta with the drive. Since the pulses are generated independently of the cavity, the Floquet scattering matrix becomes $S_n(E)= t(E_n) \mathcal J_n$, which includes the transmission amplitude for an excited electron to reach the drain. 

\emph{Electric current.} The electric current now reads~\cite{Moskalets_book}
\begin{equation}
I(t)=\frac{\mathrm{e}}{h}\int\limits_{-\infty}^\infty d E\sum_{m,n}S_{n}^\ast(E)S_{m}(E)\mathcal{F}_n(E) e^{i\Omega t (n-m)},
\label{eq:FloquetCurrent}
\end{equation}
where $\mathcal{F}_n(E) =f_S(E)-f_D(E_n)$ is the difference between the Fermi functions in the source and the drain. The electrodes have the same temperature $T$ and chemical potential $\mu=0$, such that $f_S(E)=f_D(E)=f(E)$. 

To evaluate the current, we consider the product
\begin{equation}
\label{eq:Sclint}
S_{n}^\ast(E)S_{m}(E)=\mathcal J_n^\ast \mathcal J_m T_1[T_{\mathrm{cl}}+T_{\mathrm{int}}(E)]T_2 e^{i\Omega\tau_0 (n-m)},
\end{equation}
which we have split into two terms by defining
\begin{equation}
T_{\mathrm{cl}}=\sum_{k=0}^\infty \left[R_2R_1e^{i2\Omega \tau_0 (n-m) }\right]^{k},
\end{equation}
where $T_{1,2}=t_{1,2}^2$ and $R_{1,2}=r_{1,2}^2$, and
\begin{equation}
T_{\mathrm{int}}(E)=\sum^\infty_{k\neq l} (r_2r_1)^{k+l}e^{i2\Omega\tau_0 (ml-nk)}e^{i2E\tau_0 (l-k)/\hbar}.
\end{equation}
The current corresponding to the first term then becomes $I_{\mathrm{cl}}(t)=T_1 T_2\sum_{k=0}^\infty (R_2R_1)^{k}I_{\mathrm{in}}(t-\tau_{k})$,
where $I_{\mathrm{in}}(t)=(\mathrm{e}^2/h)V(t)$ is the current from the source, and  $\tau_{k}=(2k+1)\tau_0$ is the time it takes to complete~$k+1/2$ roundtrips inside the cavity.  Here, $T_{\mathrm{cl}}$ is energy independent, such that the integral in Eq.~(\ref{eq:FloquetCurrent}) becomes $\int_{-\infty}^\infty d E \mathcal{F}_n(E) =n\hbar\Omega$. We can then perform the sums as $\sum_{n}  n\hbar\Omega \mathcal J_n e^{-i\Omega tn}= -i\hbar \partial_t \mathcal J(t)$
and $\sum_{m} \mathcal J_{m} e^{-i\Omega t m}=\mathcal J(t)$ and use that $i\hbar [\partial_t \mathcal J(t)]^\ast \mathcal J(t)=\mathrm{e}V(t)$. 

Evaluating the current corresponding to the second term is more involved, since $T_{\mathrm{int}}(E)$ is energy-dependent. However, using the integral in Eq.~(\ref{eq:elecInt}), we eventually find the total current, which becomes
\begin{widetext}
\begin{equation}
	I(t)=T_1 T_2\sum_{k=0}^\infty (R_2 R_1)^{k}\left\{(\mathrm{e}^2/h)V(t-\tau_{k})+2\mathrm{e}\sum_{l=1}^\infty(r_2 r_1)^{l}\chi(\tau_{l-1/2}/\tau_\beta)  \mathrm{Re}\left[G^{(1)}(t-\tau_{k}, t-\tau_{k+l})\right]\right\},
\label{eq:elec_current_final}
\end{equation}
\end{widetext}
where the temperature enters the function $\chi(x)=x/\sinh(x)$ through the thermal time scale $\tau_\beta= \hbar/\pi k_BT$~\cite{breuer:2002,Song:2012}. This time scale determines how long particles can travel before the interference is lost. We have also introduced the correlation function~\cite{Moskalets:2015}
\begin{equation}
	G^{(1)}(t, t')=\frac{\exp\{-i\phi(t,t')\}-1}{2\pi i (t-t')},
\end{equation}
having used that $\mathcal J^\ast(t)\mathcal J(t')=\exp\{-i\phi(t,t')\}$ with the phase factor $\phi(t, t')=\phi(t')-\phi(t)=\mathrm{e}\int_{t}^{t'} ds V(s)/\hbar$.

\begin{figure*}
    \centering
\includegraphics[width = 0.95\textwidth]{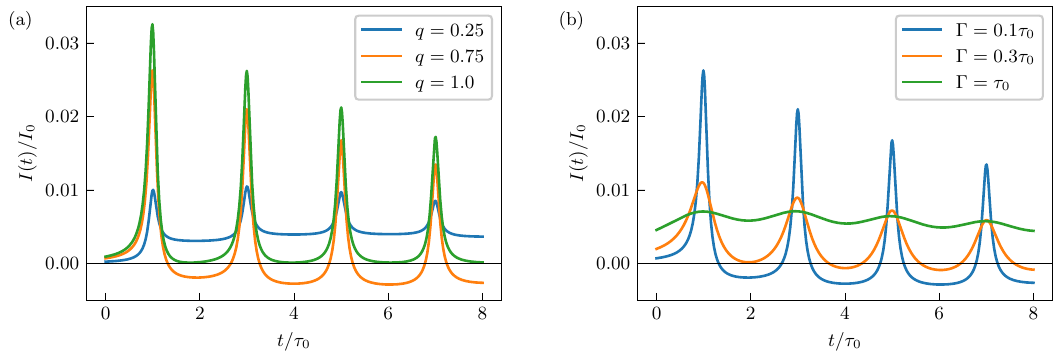}
\caption{Negative electric currents. (a) \revision{Time-dependent electric current corresponding to a Lorentzian voltage pulse with halfwidth $\Gamma = 0.1\tau_0$ and average charge of $q=0.25, 0.75$ and 1. The temperature is zero, and we have defined  $I_0 = \mathrm{e}/\tau_0$. (b)~Electric current at zero temperature for different pulse widths of $\Gamma = 0.1\tau_0, 0.3\tau_0$, and $1.0\tau_0$ and average charge of $q = 0.75$.}}   
\label{fig:electriccurrent}
\end{figure*}

Equation (\ref{eq:elec_current_final}) constitutes a central result of this Letter, and it captures the full time dependence of the electric current at any temperature and for any shape of the voltage pulses. At high temperatures, where  $\tau_\beta\ll \tau_0$, we have $\chi\simeq 0$, such that the second term in the curly brackets vanishes, and we are left with the first term. This term corresponds to classical expectations, where the injected charge bounces back and forth between the quantum point contacts, before it eventually reaches the drain. By contrast, at low temperatures, where $\chi\simeq 1$, the second term becomes important. This term describes interferences between paths through the cavity with a difference of round trips given by the integer~$l$, \revision{and it can make the current turn negative. Our theory can be experimentally tested since the temperature dependence of the interference current is fully encoded in the function~$\chi$. Thus, by increasing the temperature, one may check that the current is suppressed as expected from Eq.~(\ref{eq:elec_current_final}).}

\emph{Negative electric currents.} The period of the drive does not explicitly appear in Eq.~(\ref{eq:elec_current_final}), and we can take it to be so long that we can evaluate the current in response to just a single pulse. \revision{Thus, in Figs.~\ref{fig:setup}(b)  and~\ref{fig:electriccurrent}, we show the electric current corresponding to a single Lorentzian voltage pulse by taking $\mathcal{T}\gg \Gamma, N\tau_0$ in Eq.~(\ref{eq:levitons}), where $N$ is the largest relevant number of roundtrips.} In both figures, the electric current turns negative at certain times, and in Fig.~\ref{fig:setup}(b), we see how the negative current goes away as the temperature is increased, which destroys the interference. In Fig.~\ref{fig:electriccurrent}(a), \revision{we show the electric current for different average charges of the pulses, and we see that the current does not turn negative for an average charge of exactly one.  In Fig.~\ref{fig:electriccurrent}(b), we show the electric current for different pulse widths, and we see that the pulses have to be narrow compared with the size of the cavity to generate negative currents.}

\emph{Experimental considerations.} To provide realistic estimates for observing the negative currents, we note that a temperature of $T=100$~mK is reachable in state-of-the-art experiments, corresponding to a thermal time scale of about $\tau_\beta\simeq 20$~ps. \revision{Taking a Fermi velocity of $v_F=10^{6}$~ m$\cdot$s$^{-1}$~\cite{Roussely2018} and a distance of $L=500$~nm between the quantum point contacts, we find $\tau_0=0.5$~ps~$\ll \tau_\beta$, which is short enough to maintain phase coherence.} These numbers can be further improved by \revision{reducing the size of the cavity}, so that a quantum dot effectively is \revision{formed}~\cite{Kloss:2025}. Of course, the interference effect might be destroyed by other mechanisms than a finite electronic temperature, \revision{such as interactions and disorder}, and an observation of negative currents may require a highly controlled and ultraclean sample. \revision{For this purpose, graphene appears to be a promising host material, since recent experiments with voltage pulses have shown that interaction effects and other sources of decoherence are  suppressed at low temperatures~\cite{Assouline:2023,Chakraborti2025}.
In any case, if negative currents are measured in an experiment, they are caused by
interferences between different scattering~paths. To detect the negative currents, one may use a quantum dot to collect the transmitted charge over many periods of the drive and read it out using a sensitive charge detector.}

\emph{Heat current.} 
One might think that the negative currents arise because electrons and holes in a pulse propagate differently through the interferometer. Indeed, the negative currents do not occur for Lorentzian pulses with integer charge, where only electrons are excited. However, as we now will see, also the heat current may turn negative, although electrons and holes should carry the same amount of heat. In addition, the heat current may turn negative for Lorentzian pulses with integer charge.

\begin{figure*}
    \centering
    \includegraphics[width = 0.95\textwidth]{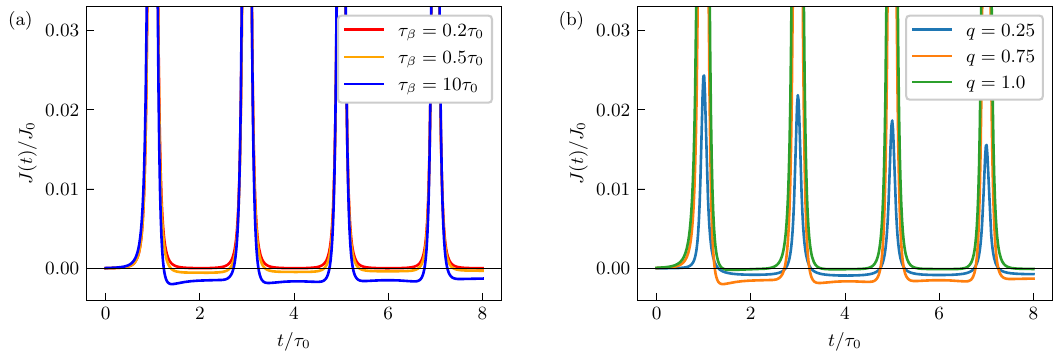}
\caption{Negative heat currents. (a) Time-dependent heat current for different temperatures given by the thermal timescale~$\tau_\beta= \hbar/\pi k_BT$. The Lorentzian voltage pulses have the halfwidth $\Gamma = 0.1\tau_0$ and an average charge of \revision{$q = 0.75$} electrons. The period of the drive is so
large that we can consider the effect of just a single pulse, and we have defined  $J_0 = I_0 V_0$ with  $V_0 = h/\mathrm{e}\tau_0$ and $I_0 = \mathrm{e}/\tau_0$. (b) Heat current at zero temperature for different pulses with an average charge of $q$.}
    \label{fig:heatcurrent}
\end{figure*}

The heat current can be expressed as~\cite{Moskalets:2004,Moskalets:2016}
\begin{equation}
J(t)\!=\!\sum_{n,m}\!\int\limits_{-\infty}^{\infty}\!\!\frac{dE}{h} E_{\frac{m+n}{2}}S_{n}^\ast(E)S_{m}(E) \mathcal F_{\frac{m+n}{2}}(E)e^{-i(m-n)\Omega t},
\label{eq:heatcurrent}
\end{equation}
which again can be evaluated based on Eq.~(\ref{eq:Sclint}). Since $T_\mathrm{cl}$ is energy independent, the corresponding heat current becomes $J_{\textnormal{cl}}(t)=T_1T_2\sum_{k=0}^\infty(R_2 R_1)^{k} J_{\textnormal{in}}(t-\tau_{k})$, where $J_{\textnormal{in}}(t) = (\mathrm{e}^2/2h)V^2(t)=I_{\mathrm{in}}(t)V(t)/2$ is the heat current that would run without the interferometer~\cite{Ludovico:2014}. Evaluating the heat current corresponding to $T_\mathrm{int}(E)$ is again involved because of the energy dependence. However, using the integrals in Eqs.~(\ref{eq:elecInt}) and  (\ref{eq:heatInt}),  the total heat current becomes
\begin{widetext}
\begin{equation}
    \begin{split}
J(t)=&T_1T_2\sum_{k=0}^\infty(R_2 R_1)^{k}\left\{ (\mathrm{e}^2/2h)V^2(t-\tau_{k})+2\hbar\sum_{l=1}^\infty(r_2 r_1)^{l}\frac{\gamma(\tau_{l-1/2}/\tau_\beta)}{\tau_l}\mathrm{Im}\left[G^{(1)}(t-\tau_k, t-\tau_{k+l})\right]\right\}\\
&+\mathrm{e}T_1T_2\sum_{k=0}^\infty\sum_{l\neq k}^\infty (r_2 r_1)^{l+k} \chi(\tau_{l-k-1/2}/\tau_\beta)\mathrm{Re}\left[G^{(1)}(t-\tau_k, t-\tau_l)\right]\left[V(t-\tau_l)+V(t-\tau_k)\right]/2,
\end{split}
\end{equation}
\end{widetext}
where we have defined the function $\gamma(x)= \chi^2(x)\cosh(x)$.

\emph{Negative heat currents.} In Fig.~\ref{fig:heatcurrent}, we show the  heat current for a single Lorentzian voltage pulse. In both panels, the heat  current turns negative at certain times, and in Fig.~\ref{fig:heatcurrent}(a), we see how the negative currents go away as the temperature is increased. In Fig.~\ref{fig:heatcurrent}(b), we show the heat current for different  average charges carried by the pulses, and we see that the heat current also turns slightly negative  for pulses with integer charge that carry no holes~\cite{Moskalets:2016,Ludovico:2014}. As such, the negative currents do not seem to occur because electrons and holes propagate differently through the cavity. Rather, they arise because of interferences between scattering paths through the cavity with a different number of round trips.

\emph{Other pulses.} So far, we have focused on Lorentzian pulses; however, our analytic expressions apply equally well to other voltage pulses. As an example, we show results in the Appendix for Gaussian pulses, which were considered in the earlier numerical simulations~\cite{Gaury:2014,Kloss:2025}. For the Gaussian pulses, we also find negative electric currents and negative heat currents. Moreover, for the electric current, our analytic expression in Eq.~(\ref{eq:elec_current_final}) agrees well with the numerical simulations of Ref.~\cite{Gaury:2014}.

\emph{Conclusions.} \revisiontwo{We have presented an analytic theory of negative currents in Fabry-Pérot cavities, showing that they indeed are rooted in interfering scattering paths through the device with different numbers of roundtrips.} Our theory can be tested in future experiments as it predicts how the effect should be washed out with increasing temperature. We have also shown that a similar phenomenon is expected for the heat current, which may also turn negative. Here, we have considered a Fabry-Pérot cavity made up of two quantum point contacts, however, our predictions also apply to other similar interferometers, for instance, a normal-metal/insulator/superconductor junction, where the insulating layer and the superconductor both function as reflecting mirrors~\cite{Roussel:2025}. Our work also has implications for the field of quantum thermodynamics regarding the currents that flow and the work that is performed because of an external driving field~\cite{Vinjanampathy:2016}.

\emph{Acknowledgments.} We thank P.~Burset, M.~Kari, J.~Kotilahti, M.~Moskalets, and B.~Roussel for useful discussions and acknowledge the support from the Finnish Doctoral School in Quantum Science and Technology and Research Council of Finland through the Finnish Centre of Excellence in Quantum Technology (grant number 352925) and the Finnish Quantum Flagship. 

\begin{figure*}
    \centering
    \includegraphics[width = 0.95\textwidth]{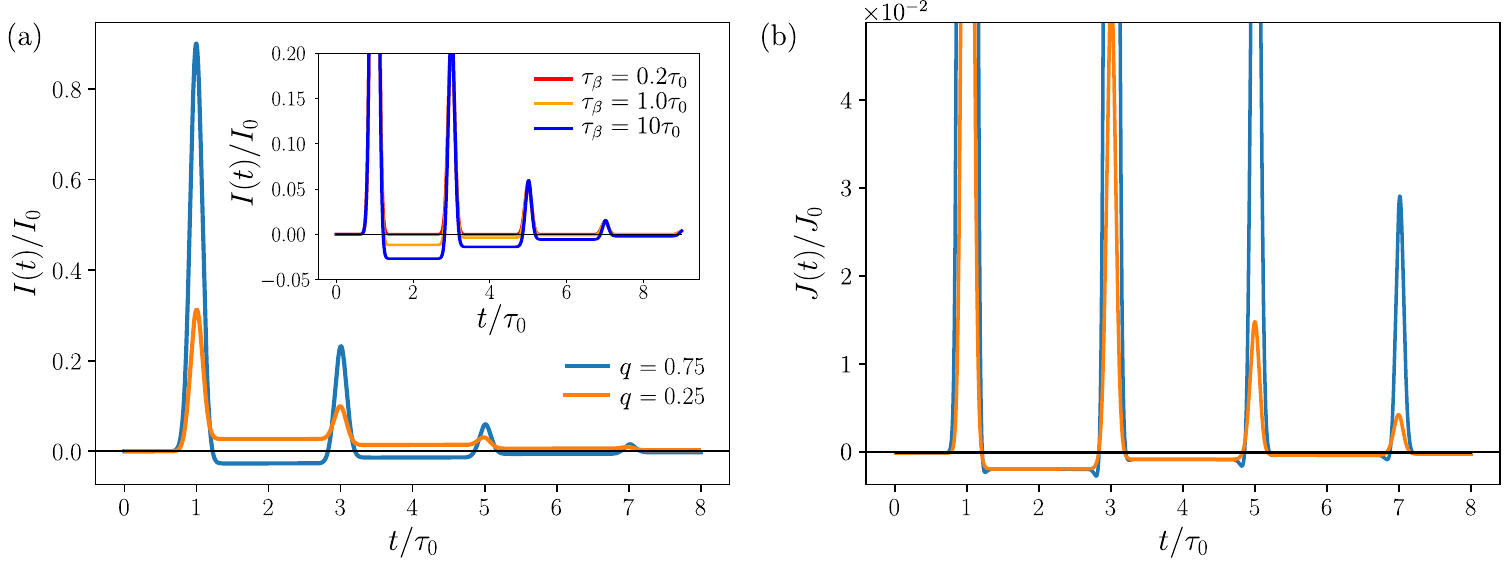}
    \caption{Negative currents for Gaussian pulses. (a) Electric current at zero temperature for the Gaussian pulse in Eq.~(\ref{eq:Vgaussian}) with $\Gamma = 0.1\tau_0$ and charge $q$. \revision{The reflection probabilities are~$R_{1,2}=0.5$.} The inset shows the electric current for different temperatures given by the thermal timescale $\tau_\beta= \hbar/\pi k_BT$ for pulses with a charge of $q = 0.75$ electrons. (b) Time-dependent heat current corresponding to the results in panel (a). Here, we have defined $I_0 = \mathrm{e}/\tau_0$ and $J_0 =I_0 V_0$ with $V_0 = h/\mathrm{e}\tau_0$.}
\label{fig:gaussianpulses}
\end{figure*}

\appendix
\emph{Appendix.}\\
\emph{Integrals.} For the electric current, we use that 
\begin{equation}
	\int_{-\infty}^\infty d x \frac{iae^{iax}}{1+e^{x}}= 
	\frac{\pi a}{\sinh{(\pi a)}}=\chi (\pi a),
	\label{eq:elecInt}
\end{equation}
where $a$ is a real constant, and we have defined the function $\chi$. For the heat current, we also need the integral
\begin{equation}
	\int_{-\infty}^\infty d x\frac{a^2 x e^{iax}}{1+e^x}=
	(\pi a)^2\frac{\cosh{(\pi a)}}{\sinh^2{(\pi a)}}=\gamma(\pi a),
	\label{eq:heatInt}
\end{equation}
having defined the function $\gamma$. Both integrals can be found using contour integration in the complex plane.

\emph{Gaussian pulses.} In the main text, we focus on Lorentzian voltage pulses, which have been realized in several experiments~\cite{Dubois:2013,Jullien:2014,Assouline:2023}. However, to connect with the numerical simulations of Ref.~\cite{Gaury:2014}, which used Gaussian voltage pulses, we here consider a pulse of the form
\begin{equation}
	V(t) = q \bar V_0 e^{-(t/t_0)^2}
	\label{eq:Vgaussian}
\end{equation}
where $q$ is the emitted charge, and we have defined the width $t_0=\Gamma/\sqrt{\ln{2}}$ and the prefactor $\bar V_0 = 2\hbar\sqrt{\pi}/\mathrm{e}t_0$.

In Fig.~\ref{fig:gaussianpulses}, we show the electric current and the heat current generated by the pulses. The results in Fig.~\ref{fig:gaussianpulses}(a) agree well with the simulations in Ref.~\cite{Gaury:2014}, and we see clear differences compared with the results in Fig.~\ref{fig:electriccurrent} for Lorentzian pulses. The inset shows how the negative currents are washed out by an increasing temperature. In Fig.~\ref{fig:gaussianpulses}(b), the heat current also turns negative.
%

\end{document}